\documentclass[12pt]{article}

\usepackage{times}

\usepackage{graphicx}

\newenvironment{sciabstract}{%
\begin{quote} \bf}
{\end{quote}}

\newcounter{lastnote}
\newenvironment{scilastnote}{%
\setcounter{lastnote}{\value{enumiv}}%
\addtocounter{lastnote}{+1}%
\begin{list}%
{\arabic{lastnote}.} {\setlength{\leftmargin}{.22in}} {\setlength{\labelsep}{.5em}}} {\end{list}}

\title{Coherent Control of a Single Electron Spin with Electric Fields}

\author
{K. C. Nowack$^{\dagger,1,\ast}$, F. H. L. Koppens$^{\dagger,1}$, Yu. V. Nazarov$^{1}$, L. M. K. Vandersypen$^{1,\ast}$ \\
\\
\normalsize{$^{1}$Kavli Institute of Nanoscience, Delft University of Technology}\\
\normalsize{ PO Box 5046, 2600 GA Delft,
The Netherlands}\\
\normalsize{$^\ast$To whom correspondence should be addressed; E-mail:}\\
\normalsize{ k.c.nowack@tudelft.nl, l.m.k.vandersypen@tudelft.nl.} }

\date{}

\begin{document}

\baselineskip24pt

\maketitle

\begin{sciabstract}
Manipulation of single spins is essential for spin-based quantum information processing. Electrical control instead of magnetic control is particularly appealing for this purpose, since electric fields are easy to generate locally on-chip. We experimentally realize coherent control of a single electron spin in a quantum dot using an oscillating electric field generated by a local gate. The electric field induces coherent transitions (Rabi oscillations) between spin-up and spin-down with $90^\circ$ rotations as fast as $\sim$55\,ns. Our analysis indicates that the electrically-induced spin transitions are mediated by the spin-orbit interaction. Taken together with the recently demonstrated coherent exchange of two neighboring spins, our results demonstrate the feasibility of fully electrical manipulation of spin qubits.
\end{sciabstract}

\renewcommand{\thefootnote}{}

\footnotetext{$^\dagger$Both authors contributed equally to this work.}

\renewcommand{\thefootnote}{\arabic{footnote}}

\newpage

 Spintronics and spin-based quantum information
processing provide the possibility to add new functionality to
today's electronic devices by using the electron spin in addition to
the electric charge~\cite{awschalom02}. In this context, a key
element is the ability to induce transitions between the spin-up and
spin-down states of a localized electron spin, and to prepare
arbitrary superpositions of these two basis states. This is commonly
accomplished by magnetic resonance, whereby bursts of a resonant
oscillating magnetic field are  applied~\cite{poole}. However,
producing strong oscillating magnetic fields in a semiconductor
device requires specially designed microwave
cavities~\cite{simovic06} or microfabricated
striplines~\cite{koppens06}, and has proven to be challenging. In
comparison, electric fields can be generated much more easily,
simply by exciting a local gate electrode. In addition, this allows
for greater spatial selectivity, which is important for local
addressing of individual spins. It would thus be highly desirable to
control the spin by means of electric fields.

Although electric fields do not couple directly to the electron
spin,  indirect coupling can still be realized by placing the spin
in a magnetic field gradient~\cite{tokura06} or in a structure with
a spatially varying $g$-tensor,
 or simply through spin-orbit interaction,  present in most semiconductor
structures~\cite{bychkov84,dresselhaus55}. Several of these
mechanisms  have been employed to electrically manipulate electron
spins in two dimensional electron systems
\cite{kato03a,kato03b,salis01,schulte05}, but proposals for coherent
electrical control at the level of a single
spin~\cite{golovach06,levitov03,tokura06,debald05,walls07} have so
far remained unrealized.

We demonstrate coherent single spin rotations  induced by an
oscillating electric field. The electron is confined in a
gate-defined quantum dot (see Fig. 1A) and we use an adjacent
quantum dot, containing one electron as well, for read-out. The ac
electric field is generated through excitation of one of the gates
that forms the dot, thereby periodically displacing the electron
wavefunction around its equilibrium position (Fig. 1B).

The experiment consists of four stages (Fig. 1C). The device is
initialised in a spin-blockade regime where two excess electrons,
one in each dot, are held fixed with parallel spins (spin triplet),
either pointing along or opposed to the external magnetic field (the
system is never blocked in the triplet state with anti-parallel
spins, because of the effect of the nuclear fields in the two dots
combined with the small interdot tunnel coupling, see
~\cite{koppens07b} for full details). Next, the two
spins are isolated by a gate voltage pulse, such that electron
tunneling between the dots or to the reservoirs is forbidden.  Then,
one of the spins is rotated by an ac voltage burst applied to the
gate, over an angle that depends on the length of the burst~\cite{SOM}
(most likely the spin in the right dot, where the
electric field is expected to be strongest). Finally, the read-out
stage allows the left electron to tunnel to the right dot if and
only if the spins are anti-parallel. Subsequent tunneling of one
electron to the right reservoir gives a contribution to the
current. This cycle is continuously repeated, and
the current flow through the device is thus proportional to the
probability of having antiparallel spins after excitation.

To demonstrate that electrical excitation can indeed induce
single-electron spin flips, we apply a microwave burst of constant
length to the right side gate and monitor the average current flow
through the quantum dots as a function of external magnetic field
$\mathbf{B_\mathrm{ext}}$ (Fig. 2A). A finite current flow is
observed around the single-electron spin resonance condition, i.e.
when $|\mathbf{B_\mathrm{ext}}| = h f_\mathrm{ac} / g
\mu_\mathrm{B}$, with $h$ Planck's constant, $f_\mathrm{ac}$ the
excitation frequency, and $\mu_\mathrm{B}$ the Bohr magneton. From
the position of the resonant peaks measured over a wide magnetic
field range (Fig. 2B) we determine a $g$-factor of $|g|=0.39 \pm
0.01$, which is in agreement with other reported values for
electrons in GaAs quantum dots~\cite{hansonrmp06}.

In addition to the external magnetic field, the electron  spin feels
an effective nuclear field $B_\mathrm{N}$ arising from the hyperfine
interaction with nuclear spins in the host material and fluctuating
in time~\cite{khaetskii02,merkulov02}. This nuclear field modifies
the electron spin resonance condition and is generally different in
the left and right dot (by $\Delta B_\mathrm{N}$).  The peaks shown
in Fig. 2A are averaged over many magnetic field sweeps and have a
width of about 10-25 mT. This is much larger than the expected
linewidth, which is only 1-2 mT given by the statistical fluctuations of
$B_\mathrm{N}$~\cite{johnson05a,koppens05}. Looking at individual
field sweeps measured at constant excitation frequency, we see that
the peaks are indeed a few mT wide (see Fig. 2C), but that the peak
positions change in time over a range of $\sim$ 20mT. Judging from
the dependence of the position and shape of the averaged peaks on
sweep direction, the origin of this large variation in the nuclear
field is most likely dynamic nuclear
polarization~\cite{baugh07,rudner06,koppens06,klauser06,laird07}.

In order to demonstrate coherent control of the spin, the length of
the microwave bursts was varied, and the current level monitored. In
Fig. 3A we plot the maximum current per magnetic field sweep as a
function of the microwave burst duration, averaged over several
sweeps (note that this is a more sensitive method than averaging
the traces first and then taking the maximum)\cite{SOM}. The maximum current
exhibits clear oscillations as a function of burst length. Fitting
with a cosine function reveals a linear scaling of the oscillation
frequency with the driving amplitude (Fig. 3B), a characteristic
feature of Rabi oscillations, and proof of coherent control of the
electron spin via electric fields.

The highest Rabi frequency we  achieved is $\sim 4.7$\,MHz (measured
at $f_\mathrm{ac}=15.2$\,GHz) corresponding to a $90^\circ$ rotation
in $\sim 55$ ns, which is only a factor of two slower than those
realized with magnetic driving~\cite{koppens06}. Stronger electrical
driving was not possible because of photon-assisted-tunneling. This
is a process whereby the electric field provides energy for one of
the following transitions: tunneling of an electron to a reservoir
or to the triplet with both electrons in the right
dot. This lifts spin-blockade, irrespective of whether the spin
resonance condition is met.

Small Rabi frequencies could be observed as well. The bottom trace
of Fig. 3A shows a Rabi oscillation with a period exceeding $1.5
\mu$s (measured at $f_\mathrm{ac}=2.6$\,GHz), corresponding to an
effective driving field of only about 0.2\,mT, ten times smaller
than the statistical fluctuations of the nuclear field. The reason
the oscillations are nevertheless visible is that the  dynamics of
the nuclear bath is slow compared to the Rabi period, resulting in a
slow power law decay of the oscillation amplitude on driving
field~\cite{koppens07}.

We next turn to the mechanism responsible for resonant transitions
between spin states. First, we exclude a magnetic origin as the
oscillating magnetic field generated upon excitation of the gate is
more than two orders of magnitude too small to produce the observed
Rabi oscillations with periods up to $\sim 220$\,ns, which requires a driving
field of about 2mT ~\cite{SOM}. Second, we have seen that there is in principle a number of ways in
which an ac electric field can cause single spin transitions. What
is required is that the oscillating electric field give rise to an
effective magnetic field, $\mathbf{B_\mathrm{eff}}(t)$, acting on
the spin, oscillating in the plane perpendicular to
$\mathbf{B_\mathrm{ext}}$, at frequency $f_\mathrm{ac} = g
\mu_\mathrm{B} |\mathbf{B_\mathrm{ext}}| / h$. The $g$-tensor
anisotropy is very small in GaAs so g-tensor modulation can be ruled
out as the driving mechanism. Furthermore, in our experiment there
is no external magnetic field  gradient applied, which could
otherwise lead to spin resonance~\cite{tokura06}. We are aware of
only two remaining possible coupling mechanisms: spin-orbit
interaction and the spatial variation of the nuclear field.

In principle, moving the wavefunction in a nuclear field gradient
can drive spin transitions ~\cite{erlingsson02,tokura06} as was
recently observed~\cite{laird07}. However, the measurement of each
Rabi oscillation took more than one hour, much longer than the time
during which the nuclear field gradient is constant ($\sim 100 \mu$s
- few s). Because this field gradient and  therefore, the
corresponding effective driving field slowly fluctuates in time
around zero, the oscillations would be strongly damped, regardless
of the driving amplitude \cite{laird07}. Possibly a (nearly) static
gradient in the nuclear spin polarization could develop due to
electron-nuclear feedback. However, such polarization would be
parallel to $\mathbf{B_\mathrm{ext}}$ and can thus not be
responsible for the observed coherent oscillations.

In contrast, spin-orbit mediated driving can induce coherent
transitions ~\cite{golovach06}, which can be understood as follows. The
spin-orbit interaction in a GaAs heterostructure is given by
$H_\mathrm{SO}=\alpha(p_x \sigma_y - p_y \sigma_x) + \beta(-p_x
\sigma_x + p_y \sigma_y)$, where $\alpha$ and $\beta$ are the Rashba
and Dresselhaus spin-orbit coefficient respectively, and $p_{x,y}$
and $\sigma_{x,y}$ are the momentum and spin operators in the $x$ and $y$
directions (along the $[100]$ and $[010]$ crystal directions
respectively). As suggested in \cite{levitov03}, the  spin-orbit
interaction can be conveniently accounted for up to the first order
in $\alpha, \beta$ by applying a (gauge) transformation, resulting
in a position-dependent correction to the external magnetic field.
This effective magnetic field, acting on the spin, is proportional
and orthogonal to the field applied:
\begin{equation}
\mathbf{B}_\mathrm{eff}(x,y) = \mathbf{n}\otimes \mathbf{B}_\mathrm{ext}; \ n_x = \frac{2m^*}{\hbar} \left(-\alpha
y -\beta x\right);\; n_y = \frac{2m^*}{\hbar} \left(\alpha x +\beta y\right);\; n_z=0 \label{eq1}
\end{equation}

An electric field $\mathbf{E}(t)$ will periodically and
adiabatically displace the electron wave function (see Fig. 1B) by
$\mathbf{x}(t)=(e l_\mathrm{dot}^2/\Delta) \mathbf{E}(t)$, so the
electron spin will feel an oscillating effective field ${\mathbf
B}_\mathrm{eff}(t)\perp {\mathbf B}_\mathrm{ext}$ through the
dependence of ${\mathbf B}_\mathrm{eff}$ on the position. The
direction of $\mathbf{n}$ can be constructed from the direction of
the electric field as shown in Fig. 4C and together with the
direction of $\mathbf{B_\mathrm{ext}}$ determines how effectively
the electric field couples to the spin. The Rashba contribution
always gives $\mathbf{n}\bot \mathbf{E}$, while for the Dresselhaus
contribution this depends on the orientation of the electric field
with respect to the crystal axis. Given the gate geometry, we expect
the dominant electric field to be along the double dot axis (see
Fig. 1A) which is here either the $[110]$ or $[1\bar{1}0]$
crystallographic direction. For these orientations, the Dresselhaus
contribution is also orthogonal to the electric field (see Fig. 4C).
This is why both contributions will give
$\mathbf{B}_{\mathrm{eff}}\neq 0$ and lead to coherent oscillations
in the present experimental geometry, where $\mathbf{E} \parallel
\mathbf{B}_\mathrm{ext}$. Note that in \cite{laird07}, a very similar gate
geometry was used, but the orientation of
$\mathbf{B}_\mathrm{ext}$ was different, and it can be
expected that $\mathbf{E} \perp \mathbf{B}_\mathrm{ext}$. In that experiment, no
coherent oscillations were observed, which is consistent with the
considerations here.

An important characteristic of spin-orbit mediated driving is the
linear dependence of the effective driving field on the external
magnetic field which follows from Eq. 1 and is predicted in
\cite{golovach06, levitov03, khaetskii01}. We aim at verifying this
dependence by measuring the Rabi frequency as a function of the
resonant excitation frequency (Fig. 4A), which is proportional to
the external magnetic field. Each point is rescaled by the estimated
applied electric field (Fig. 4B). Even at fixed output power of the
microwave source, the electric field at the dot depends on the
microwave frequency due to various resonances in the line between
the microwave source and the gate (caused by reflections at the
bonding wires and microwave components). However, we use the
photon-assisted-tunneling response as a probe for the ac voltage
drop across the interdot tunnelbarrier, which we convert into an
electric field amplitude by assuming a typical interdot distance of 100 nm.
This allows us to roughly estimate the electric field at the dot for
each frequency~\cite{SOM}. Despite the large error bars, which
predominantly result from the error made in estimating the electric
field, an overall upgoing trend is visible in Fig. 4A.

For a quantitative comparison with theory,  we extract the
spin-orbit strength in GaAs, via the expression of the effective
field $\mathbf{B_\mathrm{eff}}$ perpendicular to
$\mathbf{B_\mathrm{ext}}$ for the geometry of this experiment
\cite{golovach06}
\begin{equation} |\mathbf{B}_\mathrm{eff}(t)| = 2|\mathbf{B_\mathrm{ext}}| \frac{l_\mathrm{dot}}{l_\mathrm{SO}}
\frac{e| \mathbf{E}(t)|l_\mathrm{dot}}{\Delta},
\end{equation}
with $l_\mathrm{SO}$ the spin-orbit length (for the other
definitions see Fig. 1B). Here, $l_\mathrm{SO}^{-1}=m^*(\alpha
\mp\beta)/\hbar$  for the case with the gate symmetry axis along
$[1\bar{1}0]$ or $[110]$ respectively. Via
$f_\mathrm{Rabi}=(g\mu_\mathrm{B} |\mathbf{B_\mathrm{eff}}|)/2h $,
the confidence interval of the slope in Fig. 4A gives a spin-orbit
length of $28-37 \mu$m (with a level splitting $\Delta$ in the right
dot of 0.9 meV extracted from high bias transport measurements).
Additional uncertainty in $l_\mathrm{SO}$ is due to the estimate of
the interdot distance and the assumption of a homogenous electric
field, deformation effects of the dot potential~\cite{walls07} and
extra cubic terms in the Hamiltonian~\cite{dresselhaus55}. Still,
the extracted spin-orbit length is of the same order of magnitude as
other reported values for GaAs quantum dots~\cite{hansonrmp06}.

Both the observed trend of $\mathbf{B_\mathrm{eff}}$ with
$f_\mathrm{ac}$ and the extracted range for $l_\mathrm{SO}$ are
consistent with our supposition (by elimination of other mechanisms)
that spin transitions are mediated by spin-orbit interaction. We
note  that also for relaxation of single electron spins in which
electric field fluctuations from phonons couple to the spin, it is by now well
established that the spin-orbit interaction is dominant at fields
higher than a few 100 mT~\cite{erlingsson02, khaetskii01, golovach06, hansonrmp06}. It can thus be expected
to be dominant for coherent driving as well.

The electrically driven single spin resonance reported here,
combined with the so-called $\sqrt{{\sc SWAP}}$ gate based on the exchange interaction
between two neighbouring spins \cite{petta05}, brings all-electrical
universal control of electron spins within reach. While the $\sqrt{{\sc SWAP}}$
gate already operates on sub-nanosecond timescales, single-spin
rotations still take about one  hundred nanoseconds (the main
limitation is photon-assisted-tunneling). Faster operations could be
achieved by suppressing photon-assisted-tunneling (e.g. by
increasing the tunnel barriers or operating deeper into Coulomb
blockade), by working at still higher magnetic fields, by using
materials with stronger spin-orbit interaction or through optimized
gate designs. Furthermore, the electrical control offers the
potential for spatially selective addressing of individual spins in
a quantum dot array, since the electric field is produced by a local
gate. Finally, we note that the spin rotations were realized at
magnetic fields high enough to allow for single-shot read-out of a
single spin~\cite{elzerman04}, so that both elements can be
integrated in a single experiment.

\begin{scilastnote}

\item
We thank L. P. Kouwenhoven, C. Barthel, E. Laird, M. Flatt\'{e}, I. T. Vink and T.
Meunier for discussions; R. Schouten, B. van der Enden and R.
Roeleveld for technical assistance and J. H. Plantenberg and P. C. de Groot 
for help with the microwave components. Supported by the Dutch
Organization for Fundamental Research on Matter (FOM) and the
Netherlands Organization for Scientific Research (NWO).
\end{scilastnote}

\clearpage

\begin{figure}[b]
\begin{center}
\includegraphics[width=4in]{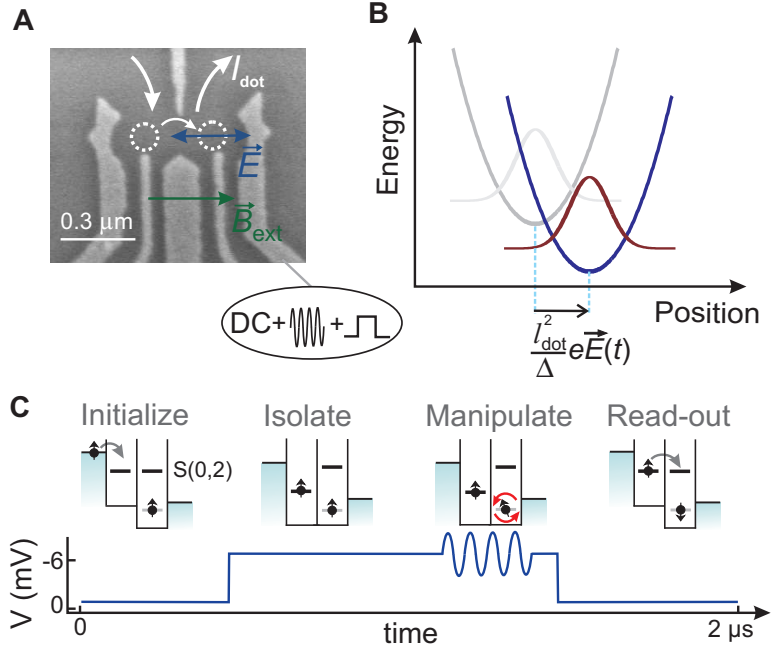}
\end{center}
\caption{(\textbf{A}) Scanning electron microscope image of a device
with the same gate structure as the one used in this experiment.
Metallic Ti\/Au gates are deposited on top of a GaAs heterostructure
which hosts a 2DEG 90 nm below the surface. Not shown is a coplanar
stripline on top of the metallic gates, separated by a dielectric
(not used in this experiment, see also \cite{koppens06}). In
addition to a dc voltage we can apply fast pulses and microwaves to
the right side gate (as indicated) through a home made bias-tee. The orientation of the in-plane external
magnetic field is as shown. (\textbf{B}) The electric field generated upon excitation
of the gate displaces the center
of the electron wavefunction along the electric field direction and
changes the potential depth. Here, $\Delta$ is the orbital energy
splitting, $l_{\mathrm{dot}}=\hbar/\sqrt{m^*\Delta}$ the size of the
dot, $m^*$ the effective electron mass, $\hbar$ the reduced Planck
constant and $\mathbf{E}(t)$ the electric field. (\textbf{C})
Schematic of the spin manipulation and detection scheme, controlled by a combination of a voltage pulse and burst, $V(t)$, applied to the right side gate. The diagrams show the double dot, with the thick black lines indicating the energy cost for adding an extra electron to the left or right dot, starting from $(0,1)$, where $(n,m)$ denotes the charge state with n and m electrons in the left and right dot. The energy cost for reaching $(1,1)$ is (nearly) independent of the spin configuration. However, for $(0,2)$, the energy cost for forming a singlet state (indicated by $S(0,2)$) is much lower than that for forming a triplet state (not shown in the diagram). This difference is exploited for initialization and detection, as explained further in the main text.}
\end{figure}

\begin{figure}[b]
\begin{center}
\includegraphics[width=4in]{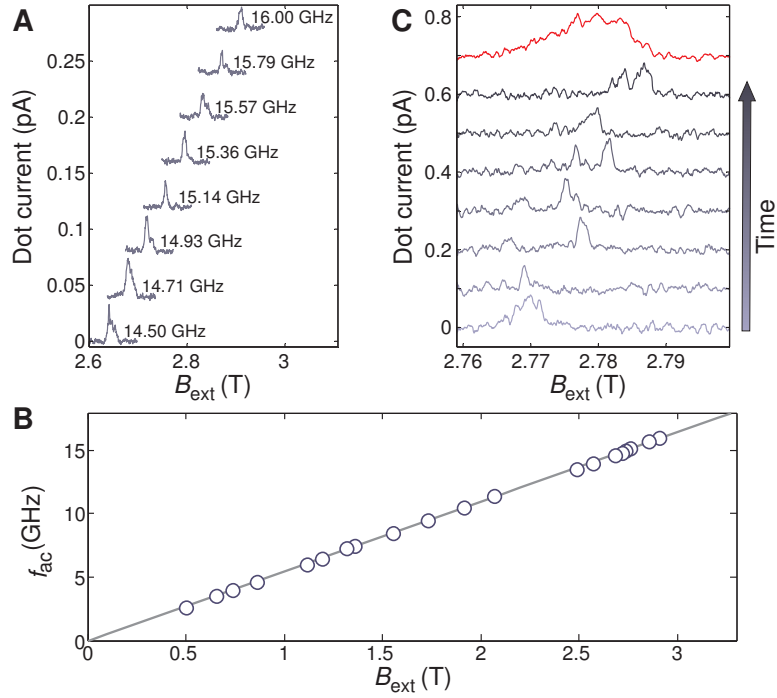}
\end{center}
\caption{(\textbf{A}) The current averaged over 40 magnetic  field
sweeps is given for eight different excitation frequencies, with a
microwave burst length of 150 ns. The traces are offset for clarity.
The microwave amplitude $V_\mathrm{mw}$ was in the range $0.9-2.2$ mV depending on
the frequency (estimated from the output power of the microwave
source and taking into account the attenuation of the coaxial lines
and the switching circuit used to create microwave bursts).
(\textbf{B}) Position of the resonant response over wider frequency
and field ranges. Errorbars are smaller than the size of the
circles.  (\textbf{C}) Individual magnetic field sweeps at
$f_\mathrm{ac}=15.2$\,GHz measured by sweeping from high to low
magnetic field with a rate of 50 mT/minute. The traces are offset by
0.1 pA each for clarity. The red trace is an average over 40 sweeps,
including the ones shown and scaled up by a factor of 5.}
\end{figure}

\begin{figure}[b]
\begin{center}
\includegraphics[width=4in]{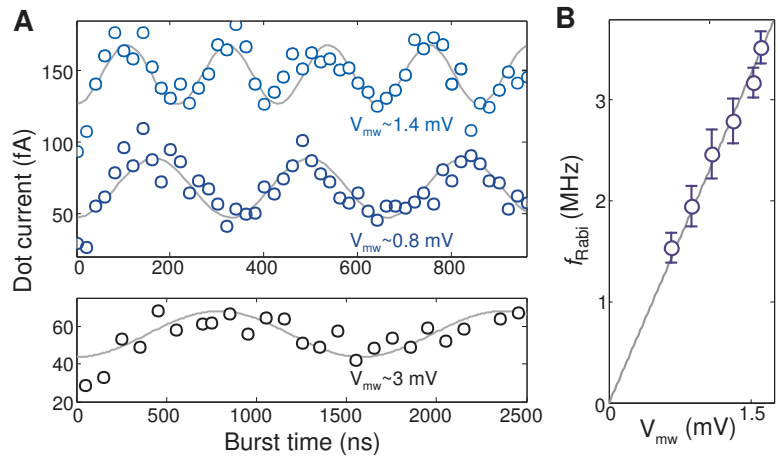}
\end{center}
\caption{(\textbf{A}) Rabi oscillations at 15.2\,GHz (blue,  average
over 5 sweeps) and 2.6\,GHz (black, average over 6 sweeps). The two
oscillations at 15.2\,GHz are measured at different amplitude of the
microwaves $V_{\mathrm{mw}}$ leading to different Rabi frequencies.
(\textbf{B}) Linear dependence of the Rabi
frequency on applied microwave amplitude measured at
$f_\mathrm{ac}=14$\,GHz. }
\end{figure}

\begin{figure}[b]
\begin{center}
\includegraphics[width=4in]{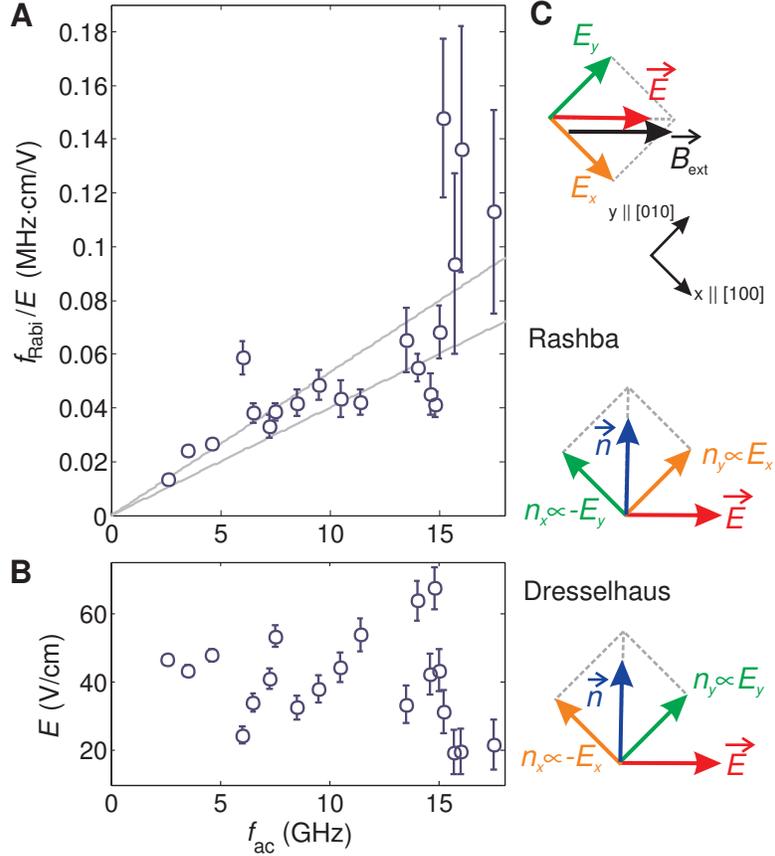}
\end{center}
\caption{(\textbf{A}) Rabi frequency rescaled with the applied
electric field for different excitation frequencies. The errorbars
 are given by
 $f_{\mathrm{Rabi}}/E\cdot\sqrt{(\delta E/E)^2+(\delta f_{\mathrm{Rabi}}/f_{\mathrm{Rabi}})^2}$ where $\delta f_{\mathrm{Rabi}}$ and $\delta E$ are the error in the Rabi frequency and electric field amplitude respectively. The grey lines
are the 95\% confidence bounds for a linear fit through the data
(weighting the datapoints by the inverse error squared).
(\textbf{B}) Estimated electric field amplitudes at which the Rabi
oscillations of \textbf(A) were measured at the respective excitation
frequencies~\cite{SOM}. (\textbf{C}) Construction of the direction of
$\mathbf{n}$ resulting from the Rashba and Dresselhaus spin-orbit
interaction for an electric field along $[110]$ following equation 1.
The coordinate system is set to the crystallographic axis $[100]$
and $[010]$.}
\end{figure}

\clearpage

\newpage
\textbf{\large{ Supporting Online Material}}
\\

\vspace{0.5cm}

\begin{description}
	\item[A] Supplementary Materials and Methods
	\item[B] Supplementary Text
	
\begin{itemize}
	\item[B.1] Extraction of Rabi oscillations from magnetic field sweeps
	\item[B.2] Estimate of the electric field amplitude at the dot
	\item[B.3] Upper bound on the ac magnetic field amplitude at the dot
\end{itemize}
\item[C] Supplementary Figures
\item[D] Supplementary References
\end{description}

\vspace{1.5cm}

\noindent {\large\textbf{A Supplementary Materials and Methods}}\\

The GaAs/AlGaAs heterostructure from which the sample were made was
purchased from Sumitomo Electric. The 2DEG has a mobility of $185 \times 10^3 \mathrm{cm}^2/\mathrm{Vs}$ at 77K,
 and an electron density of $4-5 \times 10^{11} \mathrm{cm}^{-2}$, measured at 30 mK with a different device
than used in the experiment.  Background charge fluctuations made the quantum dot
behaviour excessively irregular. The charge stability of the dot was improved
considerably in two ways. First, the gates were biased by +0.5 V relative to the 2DEG
during the device cool-down. Next, after the device had reached base temperature, the
reference of the voltage sources and I\/V converter (connected to the gates and the 2DEG)
were biased by +2 V. This is equivalent to a -2 V bias on both branches of the coplanar stripline (CPS), which
therefore (like a gate) reduces the 2DEG density under the CPS.
The sample used is identical to the one in reference \cite{koppens06}.

Based on transport measurements through the double dot, we can be nearly certain that there were only two electrons
present in the double dot. Note however that the addition of two extra electrons in one of the two dots
does not affect the manipulation and detection scheme.

The microwave bursts were created by sending a microwave signal generated by a Rohde \& Schwarz SMR40 source
through either a high isolation GaAs RF switch (Minicircuits ZASWA-2-50DR) for frequencies in the range of
10MHz to 4.6GHz or through two mixers in series (Marki Microwave M90540) for frequencies above 5GHz.
The switch and the mixers were gated by rectangular pulses from an arbitrary wave form generator (Tektronix AWG520).
The microwave bursts and voltage pulses generated by the marker channel of the same waveform generator were combined
(splitter Pasternack PE2064) and applied to the right side gate through a home made bias-tee
(rise time 150 ps and a RC charging time of $\gg$10ms at 77K).

The measurements were performed in a Oxford Instruments Kelvinox 400 HA
dilution refrigerator operating at a base temperature of ~38mK.\\

\noindent {\large \textbf{B Supplementary Text}}\\

\noindent \textbf{B.1 Extraction of Rabi oscillations from magnetic field sweeps}\\

In Fig. 2C we see that at large external magnetic field, the nuclear field fluctuates over a much larger range than $A/\sqrt{N}$, where $A$ is the nuclear field experienced by the electron spin when the nuclei are fully polarized and $N$ the number of nuclei overlapping with the electron wave function. This made it impossible in the experiment to record a Rabi oscillation at constant $B_{\mathrm{ext}}$. We therefore chose to sweep the external magnetic field through the resonance. We measured a few magnetic field sweeps per microwave burst length and averaged over the max (raw data shown in Fig. S1A)imum current values reached in each sweep.

However, when extracting the Rabi oscillation by looking at the absolute maximum per magnetic field sweep, it is not obvious that the correct Rabi period $T_{\mathrm{Rabi}}=2 h/(g \mu_{\mathrm{B}}B_{\mathrm{eff}})$ is found. For instance, a burst which produces a $2 \pi$ rotation at resonance, gives a tip angle different from $2 \pi$ away from resonance.

In order to illustrate the effect more fully, Fig. S1B shows a map of the probability for flipping a spin, calculated from the Rabi formula~\cite{poole} as a function of the detuning away from resonance and the microwave burst length.
When taking for each fixed burst length the maximum probability, a saw tooth like trace is obtained (Fig. S1C). Still the positions of the maxima remain roughly at burst lengths corresponding to odd multiples of $\pi$ and the distance between maxima corresponds to the Rabi period.

In addition, we note that every data pixel in Fig. S1A is integrated for about 50ms, so it presumably represents an average over a number of nuclear configurations.
This is additionally taken into account in Fig. S1D by averaging each point over a Gaussian distribution of detunings. The width of the distribution used in Fig. S1D corresponds to statistical fluctuations of the nuclear field
along the direction of the external magnetic field of $1.1$mT (at a driving field of $\sim 0.8$\,mT).
This assumes that on top of the large variation of the nuclear field, visible in Fig. 2C, which occurs on a minute time scale, the nuclear field undergoes additional statistical fluctuations on a faster time scale. Taking the maximum in Fig. S1D for each microwave burst length reveals a rather smooth Rabi oscillation (Fig. S1E) with a phase shift  \cite{koppens07}, and again with the proper Rabi period.

Presumably neither case, with and without averaging over a distribution of detunings, reflects the actual experimental situation in detail.
However in the simulation the Rabi period obtained from the periodicity of the maximum probability as a function of the burst length is \emph{independent} of the width of the gaussian distribution.

Finally, we remark that these conclusions are unchanged when considering the maximum current for each burst length (the current measures parallel spins versus anti-parallel spins) instead of the maximum probability for flipping a single spin. On this basis, we conclude that taking the maximum current value for each burst length gives us a reliable estimate of the Rabi period.
\\

\noindent \textbf{B.2 Estimate of the electric field amplitude at the dot}\\

The electric field generated at the dot by excitation of a gate is difficult to quantify exactly. While we can estimate the power that arrives at the sample holder from the output power of the microwave source and the measured attenuation in the line, the power that arrives at the gate is generally somewhat less (the coax is connected to the gate via bonding wires). In addition, it is difficult to accurately determine the conversion factor between the voltage modulation of the gate and the electric field modulation of the dot. We here estimate the voltage drop across the interdot tunnel barrier via photon-assisted-tunneling (PAT) measurements, and extract from this voltage drop a rough indication of the electric field at the dot.

The leakage current through the double quantum dot in the spin blockade regime as a function of the detuning
$\Delta_{\mathrm{LR}}$ (defined in Fig. S2A) shows at $B_{\mathrm{ext}}=0$\,T a peak at
$\Delta_{\mathrm{LR}}=0$ due to resonant transport and a tail for $\Delta_{\mathrm{LR}}>0$ due to inelastic
transport (emission of phonons) \cite{koppens05} (Fig. S2B). Excitation of the right side gate
induces an oscillating voltage drop across the tunnel barrier between the two dots, which leads to side peaks
at $\Delta_{\mathrm{LR}}=nhf_{\mathrm{ac}}, n=\pm1,\pm2,...$ away from the resonant peak (Fig. S2C).
These side peaks are due to electron tunnelling in combination with absorption or emission of an integer
number of photons, a process which is called photon-assisted-tunneling. In the limit where $hf_{\mathrm{ac}}$
is much smaller than the linewidth of the states $h\Gamma$ ($\Gamma$ is the tunnel rate) the individual
sidepeaks cannot be resolved, whereas for higher frequencies they are clearly visible (see Fig. S2D).

More quantitatively we describe PAT by following reference \cite{stoof96}. An ac voltage drop
$V(t)=V_{\mathrm{ac}}\cos{2\pi f_{\mathrm{ac}}t}$ across the interdot tunnel barrier modifies the tunnel rate
through the barrier as
$\tilde{\Gamma}(E)=\sum_{n=-\infty}^{+\infty}J_n^2(\alpha)\Gamma(E+nhf_{\mathrm{ac}})$. Here, $\Gamma(E)$ and
$\tilde{\Gamma}(E)$ are the tunnel rates at energy E with and without ac voltage, respectively;
$J_n^2(\alpha)$ is the square of the \textit{n}th order Bessel function of the first kind evaluated at
$\alpha=(eV_{\mathrm{ac}})/hf_{\mathrm{ac}}$, which describes the probability that an electron absorbs or
emits $n$ photons of energy equal to $hf_{\mathrm{ac}}$ (with $-e$ the electron charge). Fig. S1E shows the
current calculated from this model including a lorentzian broadening of the current peaks. A characteristic
of the $n$-th Bessel function $J_n(\alpha)$, important here, is that it is very small for $\alpha \ll n$
(i.e. when $e V_{ac} \ll n h f_{\mathrm{ac}}$) and starts to increase around $\alpha \approx n$, implying
that the number of side peaks is approximately $eV_{\mathrm{ac}}/hf_{\mathrm{ac}}$. This results in a linear envelope visible in Fig. S1E.

We extract $eV_{\mathrm{ac}}$ as the width of the region with non-zero current measured at fixed microwave frequency $f_{\mathrm{ac}}$ and amplitude
$V_{\mathrm{mw}}$. Instead of this width, we can take equivalently the number of side peaks times $hf_{ac}$ (this is possible at frequencies high enough such that individual side peaks are resolved). A reasonable estimate of the error made in determining $eV_{\mathrm{ac}}$ is $\pm h f_{\mathrm{ac}}$. Another method to extract
$V_{\mathrm{ac}}$ is to determine the slope of the envelope (for which a threshold current needs to be chosen) of the PAT response (see Fig. S2D). Varying the threshold gives a spread in the slope which defines the error of this method. We note that within the error bars both methods give the same result.

In order to estimate from $V_{\mathrm{ac}}$ the amplitude of the oscillating electric field at the
dot, $|E|$, we assume that this voltage drops linearly over the distance between the two dot centers (a rough
approximation), which is approximately 100 nm. This estimate is used in Fig. 4A in the main text, and in
the approximate determination of the spin orbit length. Note that the uncertainty in this estimate of the spin-orbit length only affects the overall scaling in Fig. 4A, but not the fact that there is an up-going trend.\\

\noindent \textbf{B.3 Upper bound on the ac magnetic field amplitude at the dot}\\

The oscillating gate voltage produces an oscillating electric field at the dot.  Here we determine an upper
bound on the oscillating magnetic field that is unavoidably generated as well. Since the distance from the
gate to the dot is much smaller than the wavelength (20 GHz corresponds to 1.5 cm), we do this in the
near-field approximation, where magnetic fields can only arise from currents (displacement currents or physical
currents).

An oscillating current can flow from the right side gate to
ground via the 2DEG, the coplanar stripline~\cite{koppens06}, or the neighbouring gates (all these elements are capacitively coupled to the right side gate). We first
consider the case of the stripline. The right side gate is about 100nm wide and overlaps with the coplanar
stripline over a length of about 10 $\mu$m, giving an overlap area of $\approx (1 \mu\mathrm{m})^2$. The gate
and stripline are separated by a 100 nm thick dielectric (calixerene \cite{holleitner03}, $\epsilon_r=7.1$), which results in a capacitance of 0.6 fF. For a maximum voltage of 10 mV applied to the right side
gate and a microwave frequency of 20 GHz, this gives  a maximum displacement current through this
capacitor of $\sim 1\mu$A. This is an upper bound as we neglect all other impedances in the path to ground. Even if this entire current flowed at a distance to the
dot of no more than 10 nm (whether in the form of displacement currents or physical currents), it would generate a magnetic field $B_{ac}$ of only $\approx 0.02$\,mT, more than two orders of
magnitude too small to explain the observed Rabi oscillations. In reality, the displacement current is
distributed along the length of the gate, and most of the current through the gate and stripline flows at a
distance very much greater than 10 nm from the dot, so $B_{ac}$
is still much smaller than 0.02 mT. The maximum magnetic field resulting from capacitive coupling to
the other gates and to the 2DEG is similarly negligible.

It is also instructive to compare the power that was applied to the gate for electric excitation of the spin with the power that was applied to the microfabricated stripline for magnetic excitation~\cite{koppens06}. For the shortest Rabi periods observed here (220 ns), the power that arrived at the sample holder was less than
$\approx -36$\,dBm (the output power of the microwave source minus the attenuation of the microwave components in between source and sample holder, measured at 6 GHz -- at higher frequencies, the attenuation
in the coax lines will be still higher). In order to achieve this Rabi frequency through excitation of the stripline, more than 100 times more power ($\approx -14$ dBm) was needed directly at the stripline~\cite{koppens06}.

The upper bounds we find for the oscillating magnetic field generated along with the electric field are thus
much smaller than the field needed to obtain the measured Rabi frequencies of a few MHz. We therefore exclude
magnetic fields as a possible origin for our observations.

\clearpage

\noindent {\large \textbf{C Supplementary Figures}}\\

\begin{figure}[ht]
\includegraphics[width=6in]{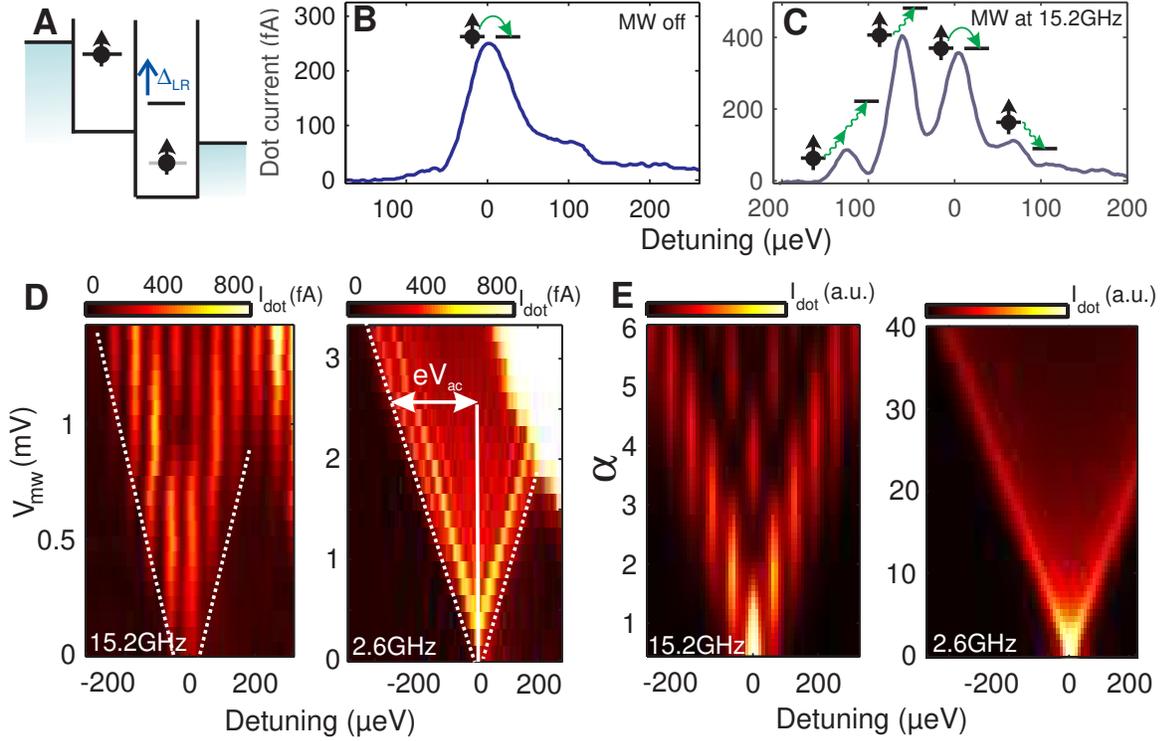}
\caption{(\textbf{A}) Magnetic field sweeps from which the topmost Rabi oscillation in Fig. 3A is extracted. The vertical axis is a combination of repeated measurements and microwave burst length (the first 5 traces correspond to a burst length of 0 ns, the following 5 to 20 ns etc.). (\textbf{B}) The simulated probability to find spin down as a function of burst length and detuning from the resonant field assuming spin up as initial state. The detuning is given in units of the driving field $B_1=B_{\mathrm{eff}}/2$ and the burst length is given in units of the Rabi period $T_{\mathrm{Rabi}}=h/(g \mu_\mathrm{B}B_1)$. (\textbf{C}) Maximum probability from (\textbf{B}) for each burst length. (\textbf{D}) Same as in (\textbf{B}) but with each pixel averaged over 75 values of the detuning, sampled from a distribution of width $\sigma$, with $\sigma=1.4 B_1$ (which corresponds to the experimental situation in \textbf{A} where $B_1\sim0.8$ mT). (\textbf{E}) Maximum probability from (\textbf{D}) for each burst length.}
\end{figure}

\begin{figure}[b]
\includegraphics[width=6in]{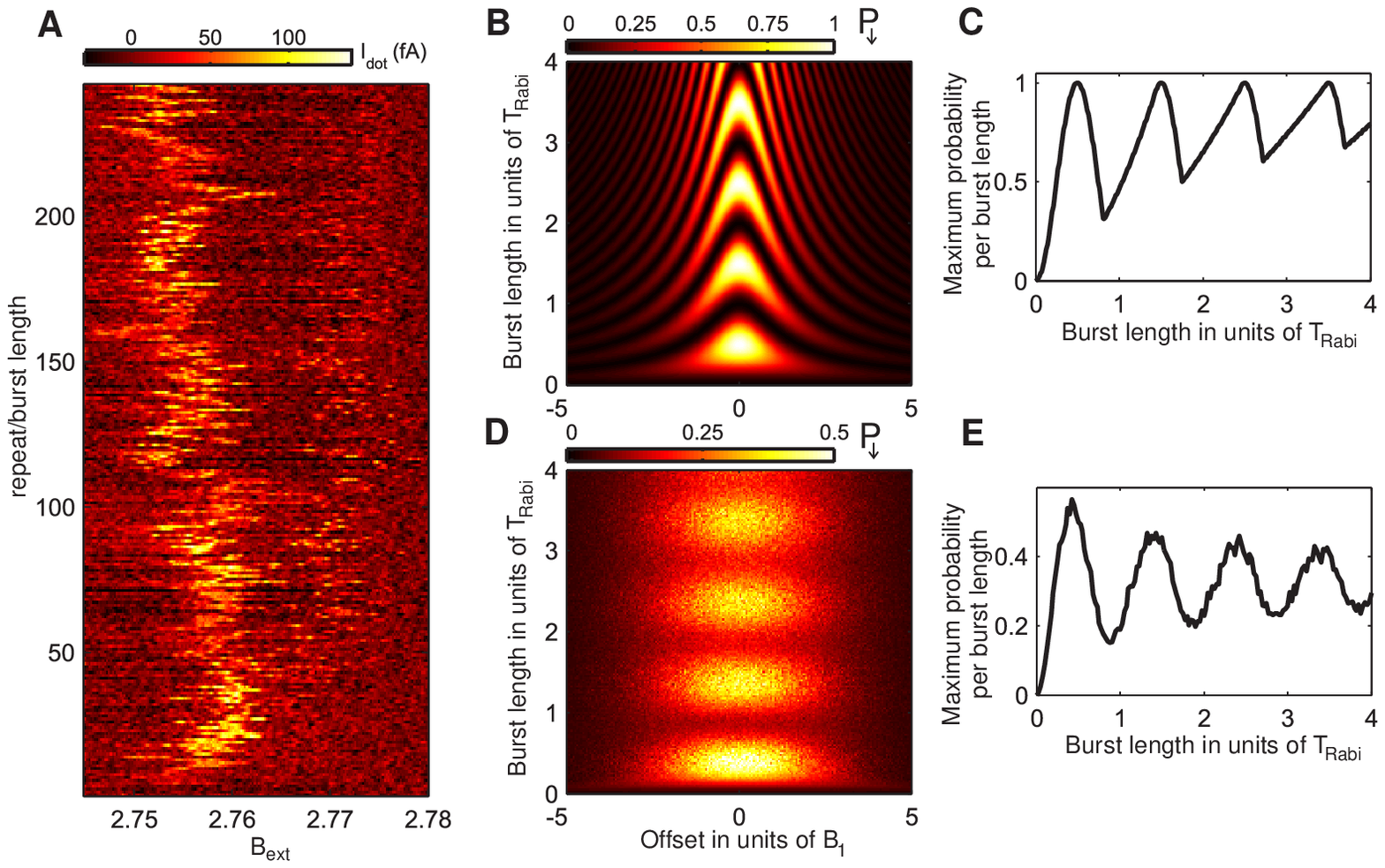}
\caption{ (\textbf{A}) Schematic of a double dot with $\Delta_{\mathrm{LR}}$ (detuning) the difference in the
energy the electron needs to access the left or right dot. (\textbf{B,C}) Current through the double dot as a
function of detuning with microwaves turned off (\textbf{B}) and on (\textbf{C}). (\textbf{D})Measured
current as a function of detuning $\Delta_{\mathrm{LR}}$ and microwave amplitude $V_{\mathrm{mw}}$ at the
gate at $f_{\mathrm{ac}}=15.2$\,GHz and 2.6\,GHz (applied in continuous wave). The external magnetic field is
zero and therefore spin blockade is lifted due to mixing of the spin states through the fluctuating nuclear
field\cite{koppens05}. At higher microwave amplitude ($V_{\mathrm{mw}}>0.5$\,mV and $1.5$\,mV respectively), the transition
to the right dot triplet state is also visible (in the upper right corner). $V_{\mathrm{mw}}$ is determined by the estimated attenuation
of the coaxial lines and the switching circuit used to create microwave bursts. (\textbf{E}) Simulated
current as a function of detuning and $\alpha=eV_{\mathrm{ac}}/(hf_{\mathrm{ac}})$ ($h$ Planck's constant)
for $f_{\mathrm{ac}}=15.2$\,GHz and 2.6\,GHz respectively. It  reproduces the linear envelope of the measured
current as well as, qualitatively, a modulation of the current amplitude in detuning. However the asymmetry with respect to
detuning visible in (\textbf{D}) as well as the observed overall increase of the current with $V_{mw}$ is not captured
in this model.  \label{SOM2}}
\end{figure}

\clearpage
\noindent {\large \textbf{D Supplementary References}}


\begin{thebibliography}{10}

\bibitem{awschalom02}
D.~Awschalom, D.~Loss, N.~Samarth, {\it {Semiconductor Spintronics and Quantum
  Computation}\/} (Springer, 2002).

\bibitem{poole}
C.~Poole, {\it Electron Spin Resonance, 2nd ed.\/} (Wiley, New York, 1983).

\bibitem{simovic06}
B.~Simovi{\v{c}} {\it et~al.\/}, {\it Review of Scientific Instruments\/} {\bf
  77}, 064702 (2006).

\bibitem{koppens06}
F.~H.~L. Koppens {\it et~al.\/}, {\it Nature\/} {\bf 442}, 766 (2006).

\bibitem{tokura06}
Y.~Tokura, W.~G. Van~der Wiel, T.~Obata, S.~Tarucha, {\it Phys. Rev. Lett.\/}
  {\bf 96}, 047202 (2006).

\bibitem{bychkov84}
Y.~A. Bychkov, E.~I. Rashba, {\it J. Phys. C\/} {\bf 17}, 6039 (1984).

\bibitem{dresselhaus55}
G.~Dresselhaus, {\it Phys. Rev.\/} {\bf 100}, 580 (1955).

\bibitem{kato03a}
Y.~Kato, R.~C. Myers, A.~C. Gossard, D.~D. Awschalom, {\it Nature\/} {\bf 427},
  50 (2003).

\bibitem{kato03b}
Y.~Kato {\it et~al.\/}, {\it Science\/} {\bf 299}, 1201 (2003).

\bibitem{salis01}
G.~Salis {\it et~al.\/}, {\it Nature\/} {\bf 414}, 619 (2001).

\bibitem{schulte05}
M.~Schulte, J.~G.~S. Lok, G.~Denninger, W.~Dietsche, {\it Phys. Rev. Lett.\/}
  {\bf 94}, 137601 (2005).

\bibitem{golovach06}
V.~N. Golovach, M.~Borhani, D.~Loss, {\it Phys. Rev. B\/} {\bf 74}, 165319
  (2006).

\bibitem{levitov03}
L.~Levitov, E.~Rashba, {\it Phys. Rev. B\/} {\bf 67}, 115324 (2003).

\bibitem{debald05}
S.~Debald, C.~Emary, {\it Phys. Rev. Lett.\/} {\bf 94}, 226803 (2005).

\bibitem{walls07}
J.~Walls, {\it http://arxiv.org/abs/0705.4231\/}  (2007).

\bibitem{koppens07b}
F.~H.~L. Koppens {\it et~al.\/}, {\it J. Appl. Phys.\/} {\bf 101}, 081706
  (2007).

\bibitem{SOM}
See supporting online material.

\bibitem{hansonrmp06}
R.~Hanson, L.~P. Kouwenhoven, J.~R. Petta, S.~Tarucha, L.~M.~K. Vandersypen,
  {\it Rev. Mod. Phys.\/} {\bf 79}, 1217 (2007).

\bibitem{khaetskii02}
A.~V. Khaetskii, D.~Loss, L.~Glazman, {\it Phys. Rev. Lett.\/} {\bf 88}, 186802
  (2002).

\bibitem{merkulov02}
I.~A. Merkulov, A.~L. Efros, M.~Rosen, {\it Phys. Rev. B\/} {\bf 65}, 205309
  (2002).

\bibitem{johnson05a}
A.~C. Johnson {\it et~al.\/}, {\it Nature\/} {\bf 435}, 925 (2005).

\bibitem{koppens05}
F.~H.~L. Koppens {\it et~al.\/}, {\it Science\/} {\bf 309}, 1346 (2005).

\bibitem{baugh07}
J.~Baugh, Y.~Kitamura, K.~Ono, S.~Tarucha, {\it Phys. Rev. Lett.\/} {\bf 99},
  096804 (2007).

\bibitem{rudner06}
M.~S. Rudner, L.~S. Levitov, {\it Phys. Rev. Lett.\/} {\bf 99}, 036602 (2007).

\bibitem{klauser06}
D.~Klauser, W.~A. Coish, D.~Loss, {\it Phys. Rev. B\/} {\bf 73}, 205302 (2006).

\bibitem{laird07}
E.~A. Laird {\it et~al.\/}, {\it http://arxiv.org/abs/0707.0557\/}  (2007).

\bibitem{koppens07}
F.~H.~L. Koppens {\it et~al.\/}, {\it Phys. Rev. Lett.\/} {\bf 99}, 106803
  (2007).

\bibitem{erlingsson02}
S.~I. Erlingsson, Y.~V. Nazarov, {\it Phys. Rev. B\/} {\bf 66}, 155327 (2002).

\bibitem{khaetskii01}
A.~V. Khaetskii, Y.~V. Nazarov, {\it Phys. Rev. B\/} {\bf 64}, 125316 (2001).

\bibitem{petta05}
J.~R. Petta {\it et~al.\/}, {\it Science\/} {\bf 309}, 2180 (2005).

\bibitem{elzerman04}
J.~M. Elzerman {\it et~al.\/}, {\it Nature\/} {\bf 430}, 431 (2004).

\end{thebibliography}

\begin{thebibliography}{1}

\bibitem{koppens06}
F.~H.~L. Koppens, C.~Buizert, K.-J. Tielrooij, I.~T. Vink, K.~C. Nowack,
  T.~Meunier, L.~P. Kouwenhoven and L.~M.~K. Vandersypen, {\it Nature\/} {\bf
  442}, 766 (2006).

\bibitem{poole}
C.~Poole, {\it Electron Spin Resonance, 2nd ed.\/} (Wiley, New York, 1983).

\bibitem{koppens07}
F.~H.~L. Koppens, D.~Klauser, W.~A. Coish, K.~C. Nowack, L.~P. Kouwenhoven,
  D.~Loss and L.~M.~K. Vandersypen, {\it Phys. Rev. Lett.\/} {\bf 99}, 106803
  (2007).

\bibitem{koppens05}
F.~H.~L. Koppens, J.~A. Folk, J.~M. Elzerman, R.~Hanson, L.~H.~W. van Beveren,
  I.~T. Vink, H.~P. Tranitz, W.~Wegscheider, L.~P. Kouwenhoven and L.~M.~K.
  Vandersypen, {\it Science\/} {\bf 309}, 1346 (2005).

\bibitem{stoof96}
T.~H. Stoof and Y.~V. Nazarov, {\it Phys. Rev. B\/} {\bf 53}, 1050 (1996).

\bibitem{holleitner03}
A.~Holleitner, {\it Applied Physics Letters\/} {\bf 82}, 1887 (2003).

\end{thebibliography}
\end{document}